\newcounter{saveeqn}
\newcommand{\alpheqn}{\setcounter{saveeqn}{\value{equation}}
\stepcounter{saveeqn}\setcounter{equation}{0}
\renewcommand{\theequation}
 {\mbox{\arabic{saveeqn}$\,$\alph{equation}}}}

\newcommand{\reseteqn}
{\setcounter{equation}{\value{saveeqn}}
\renewcommand{\theequation}{\arabic{equation}}}
\documentclass[twocolumn,showpacs,preprintnumbers ,amsmath,amssymb]{revtex4}
\usepackage{graphicx}
\usepackage{dcolumn}
\usepackage{bm}
\begin{document}
\title{Lateral and normal forces between patterned substrates induced by nematic fluctuations}
\author{F. Karimi Pour Haddadan}
 \altaffiliation
 [Present address: ]{Institute for Studies in Theoretical Physics and Mathematics (IPM),
 School of Physics, PO Box 19395-5531, Tehran, Iran }
\author{S. Dietrich}%
\address{\mbox{}\\Max-Planck-Institut f${\ddot u}$r Metallforschung, Heisenbergstr.\,3, D-70569 Stuttgart,
Germany,\\and\\
Institut f${\ddot u}$r Theoretische und Angewandte Physik, Universit${\ddot a}$t Stuttgart,
Pfaffenwaldring 57, D-70569 Stuttgart, Germany}
\date{October 28, 2004}
\date{\today}
\draft
\begin{abstract}
We consider a nematic liquid crystal confined by two parallel flat
substrates whose anchoring conditions vary periodically in one
lateral direction. Within the Gaussian approximation, we study the
effective forces between the patterned substrates induced by the
thermal fluctuations of the nematic director. The shear force
oscillates as function of the lateral shift between the patterns
on the lower and the upper substrates. We compare the strength of
this fluctuation-induced lateral force with the lateral van der
Waals force arising from chemically structured adsorbed
monolayers. The fluctuation-induced force in normal direction is
either repulsive or attractive, depending on the model parameters.
\end{abstract}
\maketitle
\section{introduction}
Liquid crystals are characterized by large thermal fluctuations in
their local orientational order arising from collective alignment
of the long axis of their constituent molecules~\cite{deg}. Due to
such soft anisotropy, liquid crystals tend to respond easily to
external forces. Confining geometries such as thin films, on which
most applications of liquid crystals are based, change the
fluctuation spectrum. This can cause not only structural
changes~\cite{harnau,kondrat} but also leads to
fluctuation-induced effective forces between the
substrates~\cite{ajdari}, also known as thermodynamic Casimir
effect. In correlated fluids such as liquid crystals this Casimir
force exhibits a universal power-law decay as a function of the
separation between the substrates~\cite{zihrel}. However, this
behavior is modified in the presence of other characteristic
scales in the system~\cite{karimi1,karimi2,ziherl1}. In the case
that the substrates are laterally modulated, discrete lateral
modes of thermal fluctuations are also excited. Under such
conditions, in addition to the forces acting perpendicularly to
the substrates, effective lateral forces arise~\cite{kardar,Emig2}
with potentially interesting technological applications. We study
the influence of anchoring conditions, which vary periodically in
one lateral direction, on the fluctuations of a uniformly ordered
nematic liquid crystal. Obviously, the periodicity $\zeta$ of the
substrate pattern gives rise to an oscillatory behavior for the
lateral force as a function of the lateral shift $\delta$ between
the substrates. For small inhomogeneities the lateral force is
proportional to $\sin (2\pi\delta/\zeta)$. (The analysis of
nonperiodic patterns would provide an understanding of nematic
phases exposed to chemically disordered substrates.) The present
study actually extends our previous work~\cite{karimi} where we
considered the case in which only one of two confining substrates
exhibits a chemical pattern so that there are no lateral forces.
Here, in addition to the fluctuation-induced lateral forces, we
calculate the lateral force between the patterned substrates
across the vacuum, i.e., the background van der Waals force acting
parallel to the substrates. This background force is generated by
the necessary chemical modulations providing the laterally varying
anchoring strengths.

In Sec.~\ref{II} our model and the theoretical formalism are
specified. In Sec.~\ref{III} the fluctuation-induced lateral force
is obtained. In Sec.~\ref{IV} we calculate the lateral van der
Waals force between the patterned substrates.
The results for the fluctuation-induced normal force are presented
in Sec.~\ref{V} and finally Sec.~\ref{VI} summarizes our results.
\section{System and Formalism}
\label{II} We consider a nematic liquid crystal confined by two
flat but chemically patterned substrates at a separation $d$. The
patterns on the both substrates consist of the same periodic
stripes of anchoring energies per area $W_a$ and $W_b$ along the
$x$ direction but shifted relative to each other by the length
$\delta$ (see Fig.~1). The substrates are translationally
invariant in the y direction. The stripes are considered to vary
with respect to the strength of homeotropic anchoring so that the
mean orientation of the director ${\bf n}$ is spatially
homogeneous but the thermal fluctuations vary laterally giving
rise to effective lateral forces.
\begin{figure}
\includegraphics[height=1.2\linewidth,angle=0]{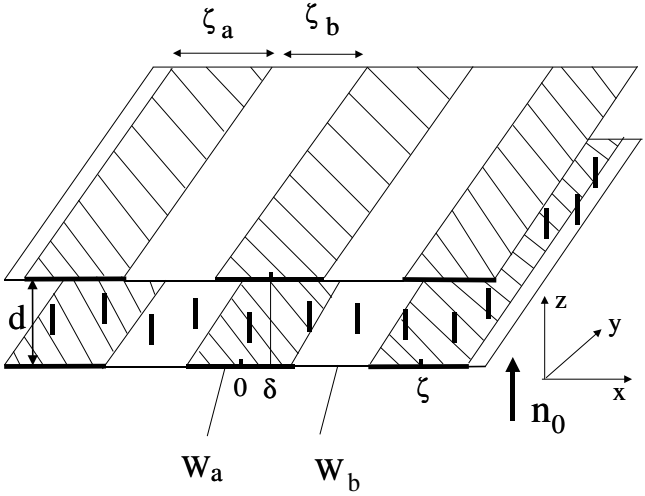}
\vspace{-3.80cm} \caption{The geometry of the nematic cell with
patterned substrates. The patterns on both substrates are the same
but shifted relative to each other. The patterns consist of
periodic stripes of anchoring energies per area $W_a$ and $W_b$
with the widths $\zeta_a$ and $\zeta_b$, respectively. The
wavelength of the periodicity is denoted as
$\zeta=\zeta_a+\zeta_b$ and the lateral shift between the origins
of the patterns on the top and the bottom substrate is denoted by
$\delta$. Anchoring at both boundaries is homeotropic everywhere
so that the thermal average of the director field ${\bf n}_0=\hat
{{\bf z}}$ is spatially homogeneous.} \label{fig0}
\end{figure}
Based on the bulk structural Frank free energy~\cite{deg} given by
\begin{eqnarray}
F_{b}[{\bf n}]=\frac{1}{2}\int_{V}{\rm
d}^3x\Big[K_1(\mbox{\boldmath $\nabla$}\cdot{\bf n})^2 +K_2({\bf
n}\cdot \mbox{\boldmath $\nabla$}\times{\bf n})^2
\nonumber\\+K_3({\bf n}\times \mbox{\boldmath $\nabla$}\times{\bf
n})^2\Big], \label{eq0}
\end{eqnarray}
where $V$ is the nematic volume, $K_1$, $K_2$, and $K_3$ are the
splay, the twist, and the bend elastic constants, respectively,
the free energy of Gaussian fluctuations in the one-constant
approximation reads
\begin{equation}
F_{\rm b}[\nu_1,\nu_2]=\sum_{i=1}^{2}{K\over 2}\int_{V} {\rm
d}^3x\left[\nabla \nu_{i}({\bf x},z)\right]^{2}, \label{eqE}
\end{equation}
where $\nu_i$, ${i=1,2}$, is either of the two independent
components of the fluctuating part $\delta \mbox{\boldmath $\nu$}
={\bf n}-{\bf n}_0$ of the director ${\bf n}$, $K=K_1=K_2=K_3$ is
the effective elastic constant, and ${\bf x}=(x,y)$ are the
lateral components of the Cartesian coordinates ${\bf r}=({\bf
x},z)$.

To describe the interaction of the liquid crystal and the
substrates, we employ the Rapini-Papoular surface free energy
given by
\begin{eqnarray}
F_{s}[{\bf n}]=-{1\over 2}\int_{S} {\rm d}^2x \,W^{z=0}({\bf
x})\,({\bf n}\cdot \hat{{\bf z}})^2-\nonumber\\{1\over 2}\int_{S}
{\rm d}^2x \,W^{z=d}({\bf x})\,({\bf n}\cdot \hat{{\bf z}})^2\
=F_s^{z=0}+F_s^{z=d} \label{eq3}
\end{eqnarray}
where $W$ is the anchoring energy per area $S$ and $\hat {\bf z}$
is the unit vector in z-direction. Here
$W^{z=0}(x)=W_{a}a(x)+W_{b}[1-a(x)]$ at the lower substrate
located at $z=0$ and $W^{z=d}(x)=W_{a}b(x)+W_{b}[1-b(x)]$ at the
upper substrate located at $z=d$ where
\begin{equation}
a(x)=\sum_{k=-\infty}^{\infty}\Theta \Big(x-k\zeta
+{\zeta\over 4}
\Big)\Theta
\Big(k\zeta+{\zeta\over 4}-x\Big)
\label{eqa}
\end{equation}
and
\begin{equation}
b(x)=a(x+\delta)
\end{equation}
describe the stripe modulations at the lower and the upper
substrates, respectively, $\Theta(x)$ is the Heaviside step
function, and $\zeta$ is the periodicity.
The stripes have the same width $\zeta_a=\zeta_b=\zeta/2$ and are
separated by sharp chemical steps. The functions $a(x)$ and $b(x)$
equal one at the regions characterized by $W_a$ and zero elsewhere
at the lower and the upper substrates, respectively. Thus the
surface free energy of the Gaussian fluctuations given by
$F_{s}[\nu_1,\nu_2]=F_{s}[\nu_{1}]+F_{s}[\nu_{2}]$ reads
\begin{eqnarray}
F_{\rm s}^{z=0}[\nu]={1\over 2}\Big[W_a\int_{S}{\rm d}^2x\,
\big[\nu ({\bf x},z=0)\big]^2
a(x) \nonumber\\
+W_b\int_{S}{\rm d}^2x\, \big[\nu ({\bf x},z=0)\big]^2
[1-a(x)]\Big]\, \label{eq3}
\end{eqnarray}
at the lower substrate and
\begin{eqnarray}
F_{\rm s}^{z=d}[\nu]={1\over 2}\Big[W_a\int_{S}{\rm d}^2x\,
\big[\nu ({\bf x},z=d)\big]^2
b(x) \nonumber\\
+W_b\int_{S}{\rm d}^2x\, \big[\nu ({\bf x},z=d)\big]^2
[1-b(x)]\Big]\, \label{eq3}
\end{eqnarray}
at the upper substrate.

Minimization of total free energy $F=F_{\rm b}[\nu_1,\nu_2]+
F_{\rm s}^{z=0}[\nu_1,\nu_2]+F_{\rm s}^{z=d}[\nu_1,\nu_2]$ leads
to two boundary conditions: \alpheqn
\begin{eqnarray}
-K\partial_z \!\!\!\!&\nu&\!\!\!\!({\bf x},z)+W_a \nu({\bf x},z)\,a(x)\nonumber\\
&+&W_b \nu({\bf x},z)\,[1-a(x)]=0, \; z=0,\label{eq:bca}
\end{eqnarray}
\begin{eqnarray}
K\partial_z \!\!\!\!&\nu&\!\!\!\!({\bf x},z)+W_a \nu({\bf x},z)\,b(x)\nonumber\\
&+&W_b \nu({\bf x},z)\,[1-b(x)]=0, \; z=d,\label{eq:bcb}
\end{eqnarray}
\reseteqn
\!\!\!\!where $\nu$ is either $\nu_1$ or $\nu_2$.
\subsection{Path integral technique}
The normalized [see, c.f., after Eq.~(\ref{eq9})] partition
function $Z$ of the fluctuating fields $\nu_{i}$, $i=1,2$, subject
to the boundary conditions given by Eqs. (\ref{eq:bca}) and
(\ref{eq:bcb}) can be calculated within the path integral approach
(see Ref.~\cite{karimi} and references therein). The boundary
conditions act as constraints which can be implemented by delta
functions. They, in turn, can be written as integral
representations by introducing two auxiliary fields localized at
$z=0$ and $z=d$, respectively~\cite{li,Gol}. After performing the
corresponding Gaussian integrals over $\nu_i$, $i=1,2$, the path
integral reduces to a Gaussian functional integral over the
auxiliary fields with a matrix kernel $M$, so that obtaining the
result for $Z$ reduces to calculating $(\det M)^{-1/2}$. For the
geometry considered here, the matrix $M$ is found to have the
following matrix elements $M_{\alpha,\beta}, \alpha,\beta=1,2$:
\begin{widetext}
\begin{eqnarray}
&M_{11}&({\bf x},{\bf x}{'})=\Big\{\Big[1+{\lambda_b-\lambda_a\over \lambda_a}a(x)\Big]
\Big[1+{\lambda_b-\lambda_a\over \lambda_a}a(x')\Big]\nonumber\\
&+&{\lambda_b (\lambda_b-\lambda_a)\over \lambda_a}
[a(x)-a(x')]\partial_z-\lambda^2_b\partial^2_z\Big\}
G({\bf x}-{\bf x'},z-z{'})\Big |_{z=z{'}=0},\nonumber\\
&M_{12}&({\bf x},{\bf x}{'})=
\Big[1+{{\lambda_b-\lambda_a}\over \lambda_a}[a(x')+b(x)]
-2\lambda_b\partial_{z'}-{{\lambda_{b}(\lambda_b-\lambda_a)}\over \lambda_a}[a(x')+b(x)]\partial_{z'}\nonumber\\
&+&\Big({{\lambda_b-\lambda_a}\over \lambda_a}\Big)^{2}a(x')
b(x)+\lambda_b^{2}\partial^{2}_{z'})\Big]
G({\bf x}-{\bf x'},z-z{'})\Big |_{z=d,z{'}=0},\nonumber\\
&M_{21}&({\bf x},{\bf x}{'})=
\Big[1+{{\lambda_b-\lambda_a}\over \lambda_a}[a(x)+b(x')]
-2\lambda_b\partial_{z}-{{\lambda_{b}(\lambda_b-\lambda_a)}\over \lambda_a}[a(x)+b(x')]\partial_{z}\nonumber\\
&+&\Big({{\lambda_b-\lambda_a}\over \lambda_a}\Big)^{2}a(x)
b(x')+\lambda_b^{2}\partial^{2}_{z})\Big]
G({\bf x}-{\bf x'},z-z{'})\Big |_{z{'}=d,z=0},\nonumber\\
&M_{22}&({\bf x},{\bf x}{'})=\Big\{\Big[1+{\lambda_b-\lambda_a\over \lambda_a}b(x)\Big]
\Big[1+{\lambda_b-\lambda_a\over \lambda_a}b(x')\Big]\nonumber\\
&+&{\lambda_b (\lambda_b-\lambda_a)\over \lambda_a}
[b(x')-b(x)]\partial_z-\lambda^2_b\partial^2_z\Big\}
G({\bf x}-{\bf x'},z-z{'})\Big |_{z=z{'}=d},\nonumber\\
\label{eq9}
\end{eqnarray}
\end{widetext}
\!\!\!where $\lambda_{a(b)}=K/W_{a(b)}$ is the so-called
extrapolation length and $G({\bf r},{\bf r}{'}) =k_BT/(4\pi
K\,|{\bf r}-{\bf r}{'}|)$ is the two-point correlation function of
the scalar field $\nu_i$ in the bulk with its statistical weight
given by $Z_{0}^{-1}\exp{\{{K\over 2k_{B}T}\int_{V}{\rm
d}^3x\,\nu_i({\bf x},z)\triangledown ^2\nu_i({\bf x},z)\}}$ where
the normalizing factor $Z_0$ is the bulk partition function and
$k_{B}T$ is the thermal energy.

In terms of the partition function $Z$, the free energy is given
by $F=-k_{B}T\ln Z=(k_{B}T/2)\ln \det M$. We note that normalizing
$Z$ by $Z_0$ amounts to subtracting the bulk free energy, so that
the free energy $F$ includes only the surface free energy and the
finite-size contribution. The surface free energy depends neither
on $d$ nor on $\delta$, so the fluctuation-induced force ${\cal
F}=-\partial F$ reads
\begin{equation}
{\cal F}=-{k_{B}T\over 2} Tr~(M^{-1}\partial M),
\label{force}
\end{equation}
where $\partial$ is either $\partial_{\delta}$ or $\partial_d$
corresponding to lateral or normal displacements giving rise to
lateral or normal forces, respectively.
\subsection{Periodic modulation}
The matrix kernel $M$ is a functional of the patterning function
$a(x)$ on the substrates and therefore calculation of the inverse
of $M$ is nontrivial. However, in systems with in-plane symmetries
one may proceed by a lateral Fourier transformation with respect
to the lateral coordinates ${\bf x}=(x,y)$. In Ref.~\cite{emig},
it is shown how the electrodynamic Casimir force can be calculated
for a periodically modulated substrate (see also
Refs.~\cite{karimi,Emig}). In this reference, the lateral
periodicity is used to transform the matrix $M$ to a
block-diagonal form in Fourier space $({\bf p},{\bf q})$
in which $M({\bf p},{\bf q})=\int \int {\rm d}^2x\;{\rm d}^2x'
M({\bf x},{\bf x}') e^{i{\bf p}\cdot {\bf x}}e^{i{\bf q}\cdot {\bf
x}'}$. Similarly, also here the matrix elements of the block
$M_j=\Big(M_{j,kl}\Big)$ with
$M_{j,kl}(p_y,q_y)=2\pi\,\mbox{\large
$\delta$}\big(p_y+q_y\big)\,B_{kl}\left({2\pi j\over L},
p_y\right)$, $j=1,\cdots ,N=L/\zeta$, are given by
\begin{equation}
B_{kl}\left({2\pi\,j\over L},p_y\right)=N_{m=k-l}\left({2\pi
j\over L}+ {2\pi\,l\over\zeta},p_y\right) \label{eqB}
\end{equation}
for $l, k\in \mathbb{Z}$, where $L$ is the lateral extension of
the system in the $x$ direction and the $N_{m}$ are $(2\times 2)$
matrices providing the following decomposition of the matrix $M$:
\begin{eqnarray}
M({\bf p},{\bf q})&=&(2\pi)^2\,\mbox{\large $\delta$}(p_y+q_y)\sum_{m=-\infty}^{\infty}N_m(p_x,p_y)
\nonumber\\&\times\mbox{\large $\delta$}&\!\!\!\! \left(p_x+q_x+
{2 \pi m\over \zeta}\right).
\label{eq2}
\end{eqnarray}

However, it is interesting to note that the matrix $M({\bf x},{\bf
x}^{'})=M({\bf x}-{\bf x}^{'},a(x),a(x^{'}))$ can also be
represented, using a more direct derivation than in
Ref.~\cite{emig}, in a form in which $M$ is diagonal~\cite{hanke}.
In view of the discrete lateral periodicity along the $x$
direction, it is suitable to express $x$ and $x^{'}$ as
\begin{equation}
x=n\zeta+s, ~~~~ x^{'}=n^{'}\zeta+s^{'},
\end{equation}
with $n, n^{'}\in \mathbb{Z}$ and $s$, $s^{'}\in
\mathbb[0,\zeta)$. Since $a(x)$ is periodic with wavelength
$\zeta$ it follows that $a(x)=a(s)$ and
$x-x^{'}=(n-n^{'})\zeta+s-s^{'}$ depends on $n$ and $n^{'}$ only
via the difference $n-n^{'}$. This property and translational
invariance along the $y$ direction imply the Fourier decomposition
\begin{eqnarray}
\hat {M}(p_x,p_y;s,s^{'})=\sum_{n,n^{'}=-\infty}^{\infty} \int
{\rm d}y \int {\rm d}y^{'}\nonumber\\
\times M(n-n^{'},y-y^{'};s,s^{'})e^{-i
p_{x}(n-n^{'})\zeta}e^{-ip_{y}(y-y^{'})} \label{M}
\end{eqnarray}
where $\hat {M}(p_x,p_y;s,s^{'})$ is diagonal. Furthermore with
$s$, $s^{'}\in \mathbb[0,\zeta)$ one can form
\begin{eqnarray}
&C_{kl}&(p_x,p_y)=\int_{0}^{\zeta}{\rm d}s\int_{0}^{\zeta}{\rm
d}s^{'}\nonumber\\
&\times& e^{2\pi i ks/\zeta} \hat {M}(p_x,p_y;s,s^{'})e^{-2\pi i
ls^{'}/\zeta}. \label{C}
\end{eqnarray}

As expected, the different representations of $M$ in terms of the
matrices $C$ or $B$ ($C$ is the Fourier transform of $\hat {M}$
and $B$ is the Fourier transform of $e^{-ip_{x}s}\hat
{M}e^{ip_{x}s^{'}}$) do not change the final result for the force
[Eq.~(\ref{force})] which is given by~\cite{emig}
\begin{eqnarray}
{\cal F}=-{k_{B}TS\over 2\pi^2}\mbox{\Large
$\int$}_{0}^{\infty}{\rm d}p_y\int_{0}^{2\pi/\zeta}{\rm
d}p_x\nonumber\\ \times{\rm tr} \big(B^{-1}(p_x,p_y)\partial
B(p_x,p_y)\big). \label{tforce}
\end{eqnarray}
Here $tr$ denotes the partial trace with respect to the indices
$k, l$ of the infinite-dimensional matrix $B$ (or $C$) and we have
taken into account the contribution of both fluctuating components
of the director field.

In the following we continue with the block-diagonal form of the
matrix kernel $M$ since in this representation the patterning
functions $a_k$ and $b_k$ are somehow simpler than their
counterparts in the diagonal form of $M$. Accordingly, the
matrices $N_m$ [Eq.~(\ref{eq2})] are given by
\begin{widetext}
\begin{eqnarray}
N_{m}&=&
\left(
\begin{array}{cc}
\phi_{m}^{aa}(0)&\phi_{m}^{ab}(d)
\\[3mm]
\phi_{m}^{ba}(d)&\phi_{m}^{bb}(0)
\end{array}
\right ) + \delta_{m,0}\left (
\begin{array}{cc}
{z_{a}^2\over 2p}-{\lambda_b^{2}p\over 2}&
{(1-\lambda_{b}p)^{2}\over 2p}e^{-pd}
\\[3mm]
\
{(1-\lambda_{b}p)^{2}\over 2p}e^{-pd}
 &{z_{b}^2\over 2p}-{\lambda_b^{2}p\over 2}
\end{array}
\right )
\nonumber\\
&+& {\lambda_b-\lambda_a\over 2\lambda_a}\delta_{m,0} \left(
\begin{array}{cc}
0&
\Big[a_0\Big({z_{b}^c\over p}-\lambda_b\Big)+b_0\Big({z_{a}^c\over
p}-\lambda_b\Big)\Big]e^{-pd}
\\[3mm]
\Big[a_0\Big({z_{b}^c\over p}-\lambda_b\Big)+b_0\Big({z_{a}^c\over
p}-\lambda_b\Big)\Big]e^{-pd}&0
\end{array}
\right )
\end{eqnarray}
\!\!\!for $m$ even, and
\begin{equation}
N_{m}= {\lambda_b-\lambda_a\over 2\lambda_a} \left (
\begin{array}{cc}
a_{m}z_{a}\Big(
{1\over p}+{1\over p_m}\Big) &\;~~~
a_m\Big({z_{b}\over p}-\lambda_b\Big)e^{-pd}+b_m\Big({z_{a}\over p_m}-\lambda_b\Big)e^{-p_{m}d}
\\[3mm]
b_m\Big({z_{a}\over p}-\lambda_b\Big)e^{-pd}+a_m\Big({z_{b}\over p_m}-\lambda_b\Big)e^{-p_{m}d}
&~~
b_{m}z_{b}\Big(
{1\over p}+{1\over p_m}\Big)
\end{array}
\right )
\end{equation}
\end{widetext}
\!\!\!\!for $m$ odd, with
$z_{a(b)}=1+{\lambda_b-\lambda_a\over \lambda_a}a_0(b_0)$,
$z_{a(b)}^c=1+{\lambda_b-\lambda_a\over 2\lambda_a}a_0(b_0)$,
${p}_m=\sqrt {(p_x+2\pi m/\zeta)^2+p_y^2}$, $p=p_{m=0}$,
$a_0=b_0=1/2$, $a_{m=odd }=(-1)^{(|m|-1)/2}/(\pi |m|)$, $b_{m=odd
}=e^{-2\pi i m\delta/\zeta}a_m$, and
\begin{equation}
\phi_m^{ab}(d)=\Big({\lambda_b-\lambda_a\over
\lambda_a}\Big)^2{\sum_{k=-\infty}^
{\infty}{}^{\!\!\!\displaystyle '}}a_{k-m}b_{-k}{e^{-p_{k}d}\over
2p_{k}},
\end{equation}
where the prime at the summation sign indicates that in the sum the terms with
even $k$ are excluded.

\section{lateral forces}
\subsection{Fluctuation-induced shear forces}
\label{III} In the limit $d\gg\zeta$ we find that the
contributions from the elements $B_{kl}$
[Eqs.~(\ref{eqB})~and~(\ref{eq2})] to the force decrease rapidly
with increasing absolute values of $k,l$, so that the expression
for the force [Eq.~(\ref{tforce})] converges already at small
orders of $M$ with $k,l=-M,\cdots,M$. Taking into account only the
elements $B_{kl}$ [Eq.~(\ref{eqB})] for $k,l=-1,0,1$, the
asymptotic behavior of the fluctuation-induced lateral force
${\cal F}_{\rm lat}=-\partial_{\delta}F$ in the limit
$|\lambda_b-\lambda_a|/\lambda_a \ll 1$ is given by
\begin{eqnarray}
{{\cal F}_{\rm lat}(d\gg \zeta)\over
k_{B}TS}={8(\lambda_b-\lambda_a)^{2}\over\pi\zeta^{3}\lambda_a^2}
e^{-2\pi d/\zeta} \sin \Big({2\pi \delta\over
\zeta}\Big)\nonumber\\\times f(d/\zeta,\lambda_b/\zeta)
+O\Big(\Big({\lambda_b-\lambda_a\over \lambda_a}\Big)^3\Big)
\label{osci}
\end{eqnarray}
with the dimensionless function
\begin{eqnarray}
f(u,v)={1\over (1+{2\pi v})^2 }\int_{0}^{\infty}{\rm d}x\,
e^{-ux}\nonumber\\\times{1-{1+2v(1-vx)x+({x\over 2\pi})^{2}(1+4\pi
v)\over (1- vx)^2} e^{2ux}\over 1-\Big({1+vx\over
1-vx}\Big)^2e^{2ux}}. \label{osci1}
\end{eqnarray}
The lateral force oscillates as function of $\delta$ reflecting
the underlying lateral periodic pattern and its magnitude decays
exponentially as function of $d/\zeta$, because
$f(\infty,v)<\infty$.

\begin{center}
\begin{figure}
\includegraphics[height=2.10\linewidth,angle=0]{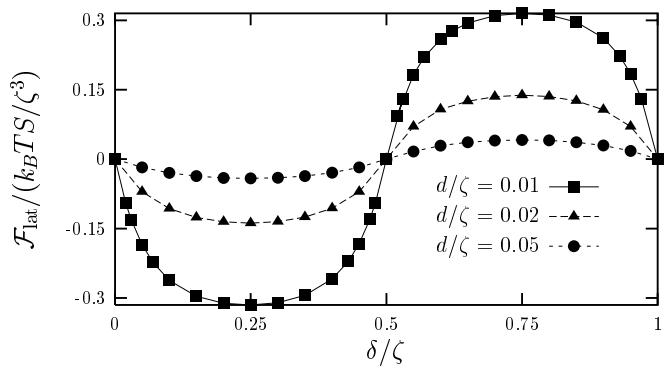}
\vspace{-12.90cm} \caption{The fluctuation-induced lateral force
${\cal F}_{\rm lat}$ in units of $k_{B}TS/\zeta^3$ between two
periodically patterned substrates at distance $d$ as function of
the shift $\delta$ in units of the periodicity $\zeta$ for
$d/\zeta=0.01,~0.02$, and $0.05$ (see Fig.~\ref{fig0}). The
anchoring on the stripes in terms of the extrapolation lengths is
taken to be $\lambda_a/\zeta=4$ and $\lambda_b/\zeta=8$. ${\cal
F}_{\rm lat}$ is antisymmetric around $\delta/\zeta=0.5$. There is
no restoring force if the misalignment is maximal, i.e., at
$\delta/\zeta=0.5$. ${\cal F}_{\rm lat}<0$ means that the plates
are pulled back towards the preferred alignment at $\delta=0$; for
${\cal F}>0$ the plates are pulled forward towards preferred
alignment at $\delta=\zeta$.} \label{fig3}
\end{figure}
\end{center}

\begin{center}
\begin{figure}
\includegraphics[height=2.050\linewidth,angle=0]{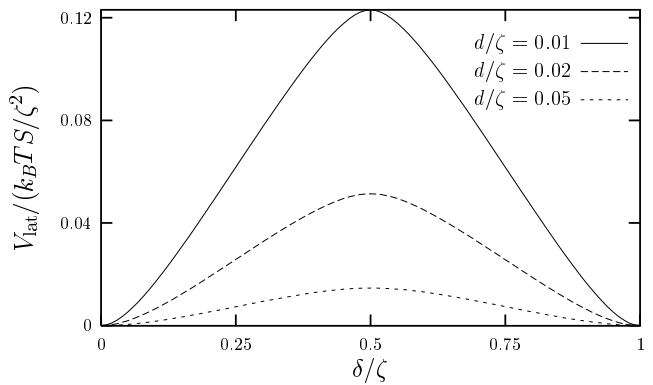}
\vspace{-12.0cm}
\caption{The effective lateral potential $V_{\rm
lat}(\delta)=-\int_{0}^{\delta}{\cal F}_{\rm lat}(\delta
^{'})d\delta ^{'}$ in units of $k_{B}TS/\zeta^2$ between two
periodically patterned substrates at distance $d$ as function of
the shift $\delta$ in units of the periodicity $\zeta$ for
$d/\zeta=0.02,~0.03$, and $0.05$ and $\lambda_a/\zeta=4$,
$\lambda_b/\zeta=8$. For all values of $d/\zeta$ the inflection
points of the potential are at $\delta/\zeta=0.25$ and $0.75$.}
\label{fig00}
\end{figure}
\end{center}

\begin{figure}
\includegraphics[height=0.6250\linewidth,angle=0]{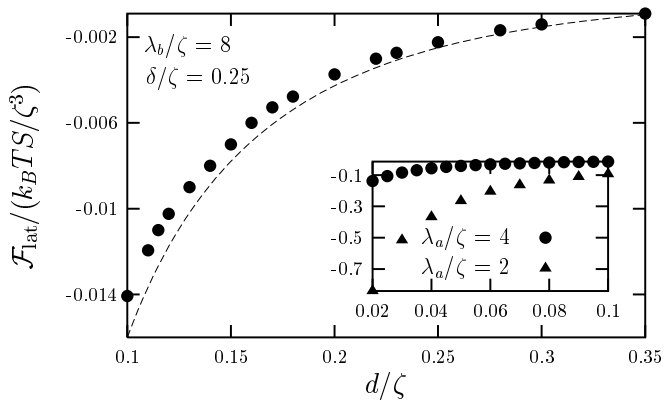}
\caption{The fluctuation-induced lateral force ${\cal F}_{\rm
lat}$ in units of $k_{B}TS/\zeta^3$ between two periodically
patterned substrates as function of the film thickness $d$ in
units of the periodicity $\zeta$. The anchoring on the stripes in
terms of the extrapolation lengths is taken to be
$\lambda_a/\zeta=4$, $\lambda_b/\zeta=8$, and the lateral shift
between the patterns is $\delta/\zeta=0.25$. Asymptotically,
${\cal F}_{\rm lat}/(k_BTS/\zeta^3)$ for $\delta/\zeta=0.25$
vanishes as $-0.003~e^{-2\pi d/\zeta}/(d/\zeta)$ (dashed line).
The variation of ${\cal F}_{\rm lat}$ for small values of
$d/\zeta$ is shown in the inset for $\lambda_a/\zeta=4$ (circles)
and $\lambda_a/\zeta=2$ (triangles). The other system parameters
remain the same. As expected the absolute value of the amplitude
of the lateral force increases upon increasing the contrast
$|\lambda_b- \lambda_a|$.} \label{fig3d}
\end{figure}

For arbitrary values of $d$, we evaluate the force in
Eq.~(\ref{tforce}) numerically. Although the matrix $B$
[Eq.~(\ref{eqB})] is infinite-dimensional, the value of the force
saturates at some finite values for $k, l$. Our numerical results
for the fluctuation-induced lateral force as function of the shift
$\delta$ for arbitrary strength of the contrast
$\lambda_b-\lambda_a$ are shown in Fig.~\ref{fig3}. The lateral
force acts against the increase of the lateral displacement
$\delta$ in the interval $[0,\zeta/2)$ by being a restoring force
and acts favorably with the increase of $\delta$ in the interval
$(\zeta/2,\zeta]$ by being a pulling force. Therefore the force is
antisymmetric with respect to $\delta/\zeta=0.5$. Upon approaching
the maximum misalignment, i.e., $\delta/\zeta=0.5$, the restoring
force vanishes. This implies that the interaction free energy
$V_{\rm lat}(\delta)=-\int_{0}^{\delta} {\cal F}_{\rm
lat}(\delta^{'})d\delta^{'}$ has its maximum at $\delta/\zeta=0.5$
where the opposing parts of the substrates face each other and
attains its minimum at $\delta=0$. The force is maximal at
$\delta/\zeta=0.25$ and $0.75$ where $V_{\rm lat}$ exhibits its
strongest dependence on $\delta$ (Fig.~\ref{fig00}). In
Fig.~\ref{fig3d} we show the decay of ${\cal F}_{\rm lat}$ as
function of $d$. Asymptotically, ${\cal F}_{\rm
lat}/(k_{B}TS/\zeta^3)$ vanishes as $-0.003~e^{-2\pi
d/\zeta}/(d/\zeta)$ for $\lambda_a/\zeta=4$, $\lambda_b/\zeta=8$,
and $\delta/\zeta=0.25$.

We note that due to the assumption $\zeta_a=\zeta_b$ interchanging
$\lambda_a$ and $\lambda_b$ leaves the system unchanged
[Fig.~\ref{fig0}]. Thus the force must be identical for
$\lambda_a\leftrightarrow \lambda_b$. While in Eqs.~(\ref{osci})
and (\ref{osci1}) this symmetry is explicitly valid up to the
second order in $(\lambda_b-\lambda_a$), the numerical results
respect this symmetry fully. This provides a very useful check of
the numerical calculations.
\subsection{Lateral van der Waals force}
\label{IV} Endowing the substrates with the envisaged stripe
patterns requires corresponding chemical patterns which in turn
involve at least two different species providing the chemical
contrast. These species do not only interact (differently) with
the nematic liquid crystal in the vicinity of the substrate,
giving rise to two different extrapolation lengths $\lambda_a$ and
$\lambda_b$, but also interact across the liquid crystal with each
other via dispersion forces. The later interaction provides a
lateral force as well which adds to the fluctuation-induced
lateral force. Note that such a lateral force due to direct
interactions is the same if the two patterned substrates are
separated by vacuum or by a nematic liquid crystal, as long as the
{\it mean} nematic order is not affected by the stripe patterns.

In order to estimate these direct interactions between the
patterned substrates, we consider each substrate to be covered by
a monolayer whose chemical composition varies periodically,
alternating between A- and B-particles. We neglect non-additivity
aspects of the dispersion forces and consider pairwise
interactions between the patterned monolayers at $z=0$ and $z=d$.
Since we consider $d$ to be large compared with the diameters of
the $A$ and $B$ particles, we can disregard that particles forming
the monolayers occupy discrete lattice sites. As pair potentials
between the two species we take Lennard-Jones potentials
\begin{equation}
u_{ij}(r)=4\epsilon_{ij}\Big[\Big({\sigma_{ij}\over r}\Big)^{12}-\Big({\sigma_{ij}\over r}\Big)^6\Big]
,~ i,j=A,B
\end{equation}
where A(B)-particles give rise to the extrapolation length
$\lambda_{a(b)}$. Since in the present context $d\gg \sigma_{ij}$,
for the lateral force only the attractive part of the pair
potentials matters. This leads to the following expression for the
van der Waals potential energy between the two monolayers:
\begin{equation}
V^{\rm vdW}_{\rm lat}(\delta)=-\int_{S} {\rm d}^{2}{\bf
x}_1\int_{S} {\rm d}^{2}{\bf x}_2
{E(x_1,x_2;\delta)\over [d^2+({\bf x}_1-{\bf x}_2)^2
]^3}
\label{23}
\end{equation}
with
\begin{eqnarray}
E(x_1,x_2;\delta)=E_{AA}a(x_1)b(x_2;\delta)+\nonumber\\
E_{BA}\big[a(x_1)-2a(x_1)b(x_2;\delta)+b(x_2;\delta)\big]\nonumber\\
+E_{BB}[1-a(x_1)][1-b(x_2;\delta)]
\end{eqnarray}
and
\begin{equation}
E_{ij}=4\epsilon_{ij}\sigma_{ij}^{6}\Sigma_i\Sigma_j
\end{equation}
where $\Sigma_{A(B)}$ is the areal number density of
$A(B)$-particles in the monolayer forming the stripe
$\lambda_{a(b)}$. Carrying out the integration over $y_1$ and
$y_2$ in Eq. (\ref{23}), one obtains
\begin{eqnarray}
V^{\rm vdW}_{\rm lat}(\delta)=-{3\pi L\over
8}\int_{-L/2}^{L/2}{\rm d}x_1\int_{-L/2}^{L/2}{\rm d}x_2
\nonumber\\
\times{E(x_1,x_2;\delta)\over [d^2+(x_1-x_2)^2]^{5/2}},
\label{pot}
\end{eqnarray}
where $L$ is the lateral extension of the system both in the $x$
and $y$ directions.
From this the lateral van der Waals force ${\cal F}_{\rm lat}^{\rm
vdW}=-\partial_{\delta}V^{\rm vdW}_{\rm lat}(\delta)$ can be
calculated:
\begin{equation}
{{\cal F}_{\rm lat}^{\rm vdW}\over S}={E\over \zeta^5}f^{\rm vdW}
(\delta/\zeta,d/\zeta) \label{eq27}
\end{equation}
with the scaling function
\begin{eqnarray}
f^{\rm vdW}(\delta/\zeta,d/\zeta)={15\pi \over
8}\sum_{n=-\infty}^{\infty}\int_{n+\delta/\zeta}^{n+1/2+\delta/\zeta}{\rm
d}X_1 \nonumber\\\times\int_{0}^{1/2}{\rm d}X_2{X_2-X_1\over
[(d/\zeta)^2+(X_2-X_1)^2]^{7/2}} \label{eq28}
\end{eqnarray}
where $E=E_{AA}-2E_{AB}+E_{BB}$. The force and the potential as
function of $\delta/\zeta$ are shown in Fig.~\ref{vdw}. The
comparison between Figs.~\ref{fig3} and \ref{vdw} (a) reveals that
the lateral van der Waals force is practically constant over a
wide range of shift values and varies steeply around the positions
of maximum and minimum misalignment while the fluctuation-induced
force varies more smoothly across all shift values. Apart from
that, the qualitative features are the same for both forces.
\begin{widetext}
\begin{center}
\begin{figure}
\includegraphics[height=1.20\linewidth,angle=0]{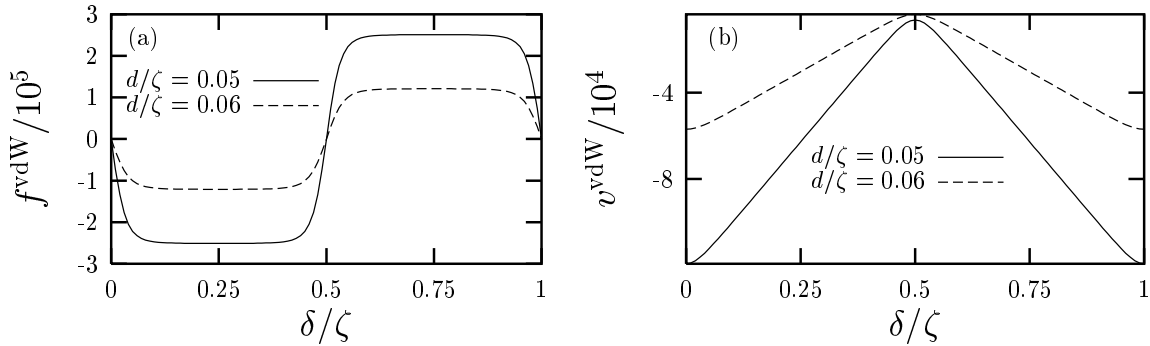}
\vspace{-16.50cm}
\caption{(a) The dimensionless scaling function $f^{\rm vdW}$
[Eqs.~(\ref{eq27})~and~(\ref{eq28})] of the lateral force induced
by direct van der Waals interactions between the chemically
patterned monolayers covering the two substrates at distance $d$
as function of the shift $\delta$ in units of the periodicity
$\zeta$ of the pattern for $d/\zeta=0.05$ and $0.06$. (b)
Corresponding lateral potential $V^{\rm vdW}_{\rm lat}(\delta)={E
S\over \zeta^4}v^{\rm vdW}_{\rm lat} [Eq.~(\ref{pot})].$}
\label{vdw}
\end{figure}
\end{center}
\end{widetext}
In order to estimate the lateral van der Waals force we assume
that the particles are closely packed within the
monolayer~\cite{lub}, so that the areal number density is
$\Sigma_{A(B)}=\sqrt{3}/(6R^2)\approx
2[\sigma_{AA(BB)}]^{-2}/\sqrt{3}$ where $R=\sigma_{AA}/2$
($\sigma_{BB}/2$) is the radius of the $A$ ($B$) particles forming
a triangular lattice. According to Table I in Ref.~\cite{getta}
typical values are $\epsilon_{ij}/k_B\approx 300~K$ and
$\sigma_{ij}\approx 0.5~{\rm nm}$. For $\zeta=200~{\rm nm}$ one
has $E_{ij}/\zeta^5\approx 1.5\times 10^{-5}~{\rm pN}/(\mu {\rm
m})^2$. According to Fig.~\ref{vdw} (a) for ${\rm d}/\zeta=0.05$
this implies ${\cal F}_{{\rm lat} ,ij}^{{\rm vdW}}\approx 3~{\rm
pNS}/(\mu {\rm m})^2$ for the contribution $\propto E_{ij}$ to the
actual lateral van der Waals force ${\cal F}_{\rm lat}^{\rm vdW}$
proportional to $E=E_{AA}-2E_{AB}+E_{BB}$. Thus for a suitably
chosen constant $E$, without compromising the goal of achieving
$\lambda_a=\lambda_b/2$ (as used in our calculations), the lateral
van der Waals force can be quite smaller than $3~{\rm pN}$ per
area $S/(\mu {\rm m})^2.$

\begin{figure}
\includegraphics[height=2.150\linewidth,angle=0]{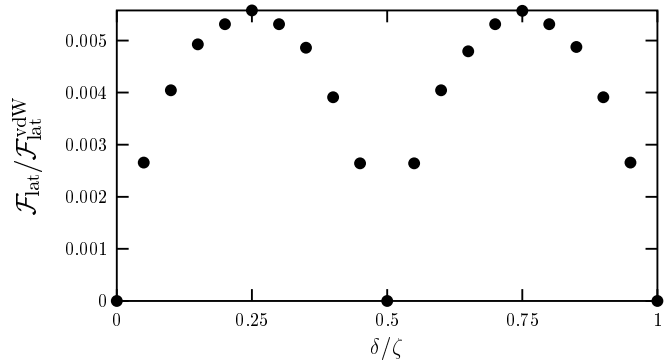}
\vspace{-13.50cm} \caption{The ratio of the fluctuation-induced
lateral force ${\cal F}_{\rm lat}$ and the lateral van der Waals
force ${\cal F}_{\rm lat}^{\rm vdW}$ as function of the shift
$\delta$ in units of the periodicity $\zeta$ for $d/\zeta=0.05$,
$\lambda_a/\zeta=4$, $\lambda_b/\zeta=8$, $E/\zeta^5\approx
1.5\times 10^{-5}~{\rm pN}/(\mu {\rm m})^2$, and at $T=290~K$.
${\cal F}_{\rm lat}$ is more prominent around $\delta/\zeta=0.25$
and $0.75$.} \label{fig56}
\end{figure}

From Fig.~\ref{fig3} one finds, for the same system parameters
$\zeta$ and $d$ considered above and at room temperature, for the
fluctuation-induced lateral force ${\cal F}_{\rm lat}\approx
0.02~{\rm pN S}/(\mu {\rm m})^2$. Thus the background lateral van
der Waals force tends to be stronger than the nematic fluctuation
induced force. However, for a suitably chosen chemical contrast of
the particles forming the chemical stripes, it appears to be
possible to determine the fluctuation-induced lateral force by
measuring the shear force once with and once without the nematic
liquid between the patterned substrates. The ratio of the two
forces for $d/\zeta=0.05$, $\lambda_a/\zeta=4$,
$\lambda_b/\zeta=8$, and $E/\zeta^5\approx 1.5\times~10^{-5}~{\rm
pN}/(\mu {\rm m})^2$ as function of $\delta$ is shown in
Fig.~\ref{fig56}. It appears that the fluctuation-induced lateral
force becomes more prominent around $\delta/\zeta=0.25, 0.75$.
From Figs.~\ref{fig3} and \ref{vdw} (a) one notices that both
${\cal F}_{\rm lat}/S$ and ${\cal F}^{\rm vdW}_{\rm lat}/S$ vanish
linearly at $\delta/\zeta=0, 0.5, 1.0$ [as
$(-0.084+0.168~\delta/\zeta)~{\rm pN}/(\mu {\rm m})^2$ and
$(-33.5+66~\delta/\zeta)~{\rm pN}/(\mu {\rm m})^2$ at
$\delta/\zeta=0.5$, respectively, for $d/\zeta=0.05$ ,
$E/\zeta^{5}=1.5\times~10^{-5}~{\rm pN}/(\mu {\rm m})^2$,
$\zeta=200~{\rm nm}$, and $k_BT=4\times 10^{-21}~J$] and since the
slope of ${\cal F}^{\rm vdW}_{\rm lat}$ is much larger than the
corresponding slope of ${\cal F}_{\rm lat}$, the ratio ${\cal
F}_{\rm lat}/{\cal F}^{\rm vdW}_{\rm lat}$ at $\delta/\zeta=0,
0.5, 1.0$ is small. From Fig.~\ref{fig56} one should not draw the
conclusion that the lateral fluctuation induced force is at most
half a percent of the corresponding lateral van der Waals
background force. This ratio is inversely proportional to $E
=E_{AA}-2E_{AB}+E_{BB}$. Figure~\ref{fig56} corresponds to a
parameter choice for which $E$ is estimated by an individual
$E_{ij}$ and not by the actual contrast expressed by E, which
vanishes for $A=B$. Accordingly, for suitable choices of $A$ and
$B$, $E$ can be significantly smaller than the $E_{ij}$ used in
Fig.~\ref{fig56} which then leads to a significantly larger ratio.
\section{fluctuation-induced normal force}
\label{V} 
\begin{figure}
\includegraphics[height=2.220\linewidth,angle=0]{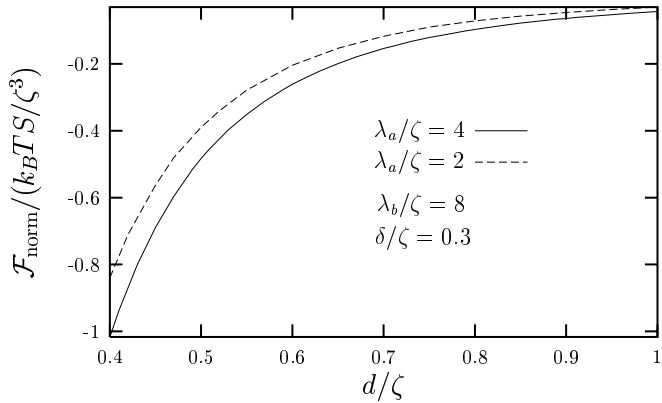}
\vspace{-19.60cm} \vspace{6cm} \caption{The fluctuation-induced
normal force ${\cal F}_{\rm norm}$ in units of $k_{B}TS/\zeta^3$
between two periodically patterned substrates as function of the
film thickness $d$ in units of the periodicity $\zeta$. The
anchoring on the stripes in terms of the extrapolation lengths is
taken to be $\lambda_b/\zeta=8$ and $\lambda_a/\zeta=4$ (full
line), $2$ (dashed line). For $d<\lambda_a,\lambda_b$ as shown
here the anchoring is weak but finite~\cite{zihrel} at both
substrates and the force is attractive. As expected the absolute
value of the amplitude of the force decreases upon decreasing the
extrapolation length. The force depends very weakly on the shift
$\delta$. Here $\delta/\zeta$ is set to $0.3$.} \label{fig1}
\end{figure}

\begin{center}
\begin{figure}
\includegraphics[height=2.220\linewidth,angle=0]{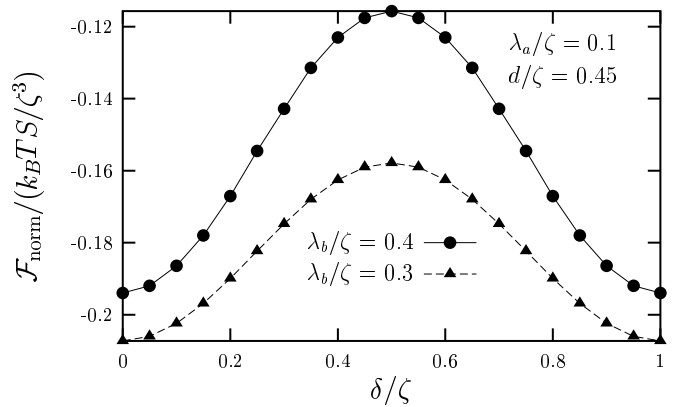}
\vspace{-19.60cm} \vspace{6cm} \caption{The fluctuation-induced
normal force ${\cal F}_{\rm norm}$ in units of $k_{B}TS/\zeta^3$
between two periodically patterned substrates at distance
$d/\zeta=0.45$ as function of the shift $\delta$ in units of the
periodicity $\zeta$. The anchoring on the stripes in terms of the
extrapolation lengths is taken to be $\lambda_a/\zeta=0.1$ and
$\lambda_b/\zeta=0.3$ (triangles), $0.4$ (circles). For
$d>\lambda_a,\lambda_b$ as shown here the anchoring is strong but
finite at both substrates and the force is attractive. As expected
the absolute value of the amplitude of the force decreases upon
increasing the extrapolation length~\cite{zihrel}. The force
oscillates as function of the shift $\delta$ and for complete
misalignment, i.e., $\delta/\zeta=0.5$ the attraction is weakest.}
\label{fig8}
\end{figure}
\end{center}

\begin{center}
\begin{figure}
\includegraphics[height=2.250\linewidth,angle=0]{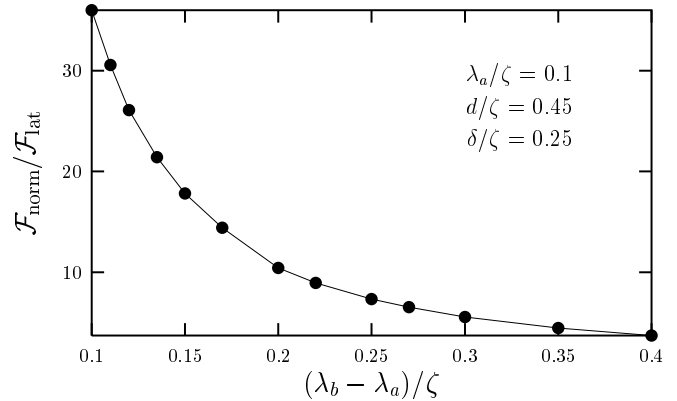}
\vspace{-13.50cm} \caption{Fluctuation-induced normal force ${\cal
F}_{\rm norm}$ divided by fluctuation-induced lateral force ${\cal
F}_{\rm lat}$ as function of the contrast $\lambda_b-\lambda_a$ in
units of the periodicity $\zeta$ for $\lambda_a/\zeta=0.1$,
$d/\zeta=0.45$, corresponding to strong but finite anchoring, and
$\delta/\zeta=0.25$. ${\cal F}_{\rm norm}$ and ${\cal F}_{\rm
lat}$ are of comparable size.}
\label{fig9}
\end{figure}
\end{center}

The effect of a periodic anchoring at one substrate,
with the second substrate being homogeneous, on the
fluctuation-induced normal force was studied in
Ref.~\cite{karimi}. It turned out that for the description of the
normal force the single patterned substrate can be replaced by a
uniform substrate with an effective anchoring strength, i.e., the
force is given by the force found between two uniform substrates
characterized by their effective anchoring. Depending on the model
parameters, the normal force is either repulsive or attractive --
corresponding to an effective similar-dissimilar or an effective
similar-similar boundary condition, respectively. In the present
case of two patterned substrates, we have calculated the normal
force numerically [Eq.~(\ref{tforce})]. Figure~\ref{fig1} shows
the dependence of the normal force ${\cal F}_{\rm norm}$ on $d$.
For $d<\lambda_a, \lambda_b$ as shown here, the force is
attractive and decays monotonically as function of $d$. In this
regime, the anchoring is weak at both substrates so that the
effect of the periodicity is not visible. We note that in this
case the fluctuation-induced normal force [Fig.~\ref{fig1}] is
about $1000~(100)$ times larger than the fluctuation-induced
lateral force [Fig.~\ref{fig3d}] for $\lambda_a/\zeta=4~(2)$.
Figure~\ref{fig8}, however, shows the behavior of ${\cal F}_{\rm
norm}$ as function of $\delta$ in the regime of strong but finite
anchoring $d>\lambda_a, \lambda_b$. In this case the normal force
is oscillatory and attractive, and its magnitude is comparable
with the fluctuation-induced lateral force [Fig.~\ref{fig9}].
\section{summary and conclusion}
\label{VI} We have calculated the fluctuation-induced forces
acting on two substrates chemically modulated with period $\zeta$
and confining a nematic film of thickness $d$ (Fig.~\ref{fig0}).
The substrates are characterized by homeotropic anchoring with
alternating extrapolation lengths $\lambda_a$ and $\lambda_b$. We
have studied the shear force as function of the lateral shift
$\delta$ between the patterns on the substrates and of their
separation $d$. For $|\lambda_b-\lambda_a|\ll \lambda_a$ and
$\zeta\ll d$, the lateral force sinusoidally oscillates as a
function of $\delta/\zeta$ and decays exponentially with $d/\zeta$
[Eq.~(\ref{osci}) and Fig.~\ref{fig3d}]. For stronger contrasts,
the lateral force and its corresponding potential have been
evaluated numerically [Figs.~\ref{fig3}~and~\ref{fig00}]. It turns
out that for a suitably chosen chemical contrast the
fluctuation-induced lateral force is comparable [Fig.~\ref{fig56}]
with the background lateral van der Waals force between the
corresponding monolayers on the top and bottom substrates forming
the chemical heterogeneity [Fig.~\ref{vdw}]. In order to complete
the picture of the forces in the presence of the patterned
substrates we have also calculated numerically the
fluctuation-induced normal forces [Figs.~\ref{fig1} and
\ref{fig8}] and found them to be comparable with the
fluctuation-induced lateral force in the case of strong anchoring
[Fig.~\ref{fig9}].

Patterning at small length scales gives rise to rich interfacial
phenomena. Surface modulations open the possibility of controlling
the morphology of wetting films and generate structural phase
transitions which are central to the behavior of structural forces
induced by distortions of the liquid crystal order parameter. In
such cases, in addition to the fluctuation-induced forces, the
substrates are subject to liquid-crystalline {\it elastic}
forces~\cite{kondrat}. Such elastic  forces scale with K
[Eq.~(\ref{eqE})] and are for large $d$ larger than the
fluctuation-induced forces which scale with $k_BT$. Since for
small $d$ the elastic forces scale as $d^{-0.5}$~\cite{harnau} but
the fluctuation-induced force as $d^{-3}$~\cite{zihrel}, the
latter can, however, even dominate for small $d$. The {\it mean
field} director contributions are completely eliminated for such
model parameters and boundary conditions for which the director
structure is uniform. In Ref.~\cite{pon} it has been demonstrated
that this uniformity can indeed occur for suitable combinations of
model parameters, even in cases of competing planar and
homeotropic anchoring conditions, which have not been considered
here. Capillary forces due to capillary condensation and the
formation of bridge phases~\cite{kocevar,holger,denis} are other
sources for the structural forces in the vicinity of the
nematic-isotropic phase transition. If the fluid is confined to
very narrow slits, patterning may give rise to capillary bridges
of different liquid crystalline order. Under such conditions, it
would be interesting to study the stress under the shear strains
by shifting the lateral substrate structures out of
phase~\cite{schoen} which might give rise to rather strong lateral
forces.

Lateral forces can technologically be used to align the parallel
substrate structures. For instance, for those ranges of the model
parameters for which the liquid crystalline lateral forces are
significant, the liquid crystal can be filled into the slit pore
to align the substrate structure and then be removed.

\end{document}